\documentclass[epj]{svjour}

\usepackage{graphicx}
\usepackage{fancyhdr}
\def\makeheadbox{{
 }}
\def\mytitle{My title} 
\def\myauthors{My name}  
\def\mytype{My type of session}
\def\mysession{My session}

\def\mytitle{Dark Matter Phenomenology of GUT-less SUSY Breaking} 
\def\myauthors{Pearl Sandick}    
\def\mytype{Contributed Talk}    
\def\mysession{Cosmology and Astrophysics}

\begin{document}
\title{Dark Matter Phenomenology of GUT-less SUSY Breaking}
\author{Pearl Sandick\inst{1}
\thanks{\emph{Email:} sandick@physics.umn.edu}%
}                     
%
%
\institute{School of Physics and Astronomy,\\
University of Minnesota, Minneapolis, MN 55455, USA}
%
\date{}
\abstract{We study models in which supersymmetry breaking appears at an intermediate
scale, $M_{in}$, below the GUT scale. That is, that the soft
supersymmetry-breaking parameters of the MSSM are universal at $M_{in}$. We demand that the lightest neutralino be the LSP, and that the relic neutralino density not conflict with measurements by WMAP
and others, and study the morphology of this constraint as the
universality scale is reduced from the GUT scale. At moderate values of $M_{in}$, we find that the allowed
regions of the $(m_{1/2},m_0)$ plane are squeezed by the requirements of
electroweak symmetry breaking and that the lightest neutralino be the LSP,
whereas the constraint on the relic density is less severe. At very low $M_{in}$,
the electroweak vacuum conditions become the dominant constraint, and a
secondary source of astrophysical cold dark matter would be necessary to
explain the measured relic density for nearly all values of the soft
SUSY-breaking parameters and $\tan{\beta}$.
} 
\maketitle
\section{Introduction}
\label{intro}

It is well known that TeV scale SUSY offers a compelling solution to
the related hierarchy and naturalness problems of the Standard Model.
It also facilitates unification of the gauge couplings, as expected in
Grand Unified Theories (GUTs) and predicts a light Higgs boson,
as favored by precision electroweak data. If R-parity is assumed to be
conserved, the lightest supersymetric particle (LSP) is stable, and,
if uncharged, is therefore a natural particle candidate
for astrophysical cold dark matter.

SUSY is, of course, broken, however the mechanism of this
breaking and how it is communicated to the observable sector are
unknown.  One option is to parametrize the breaking at a high scale with soft
SUSY-breaking mass parameters and use the renormalization group
equations (RGEs) of the low-energy effective theory to evolve them
down to lower scales. In the Constrained Minimal Supersymmetric
Standard Model (CMSSM), the soft SUSY-breaking parameters are taken to
be universal at the SUSY GUT scale, $M_{GUT} \sim 2 \times 10^{16}$ GeV. The
RGEs of the MSSM determine the weak scale observables, given five
inputs at the GUT scale, the scalar mass, $m_0$, the gaugino mass,
$m_{1/2}$, the trilinear coupling, $A_0$, the ratio of the Higgs vevs,
$\tan{\beta}$, and the sign of the Higgs mass parameter, $\mu$. In
some models it may
be more appropriate, however, to assume universality of the soft
SUSY-breaking parameters at some scale $M_{in}$ intermediate between
the GUT scale and the electroweak scale.

Here, we present the results of recent studies on the effect of lowering
the universality scale on dark matter phenomenology \cite{eos1,eos2}. We begin with a
discussion of renormalization of the soft SUSY-breaking parameters and
expectations based on simple one-loop
approximations.  Section \ref{sec:evolution} contains the core of our
results, though we limit ourselves here to the $\tan{\beta}=10$
scenario, with only a brief comment on $\tan{\beta}=50$. In Section
\ref{sec:detection} we discuss the prospects for direct detection.  Finally, conclusions are given
in Section \ref{sec:conclusions}.

\section{GUT-less Renormalization}
\label{sec:renorm}

The consequences of lowering the universality scale can be easily
understood by examining the changes to the running of the soft mass parameters at the one-loop level\footnote{Note that all
calculations carried out in making the following plots incorporate the
full two-loop RGEs.}. In the GUT-scale CMSSM, the one-loop
renormalizations of the gaugino masses are identical to those of the
corresponding gauge couplings. The gaugino masses, $M_a(Q)$, where $a=1$,
2, 3, and $Q$ is some low scale, are then proportional to $m_{1/2}$, with
the proportionality defined by the ratio of the corresponding gauge
coupling at the low scale to the coupling at the GUT scale.  The
running of the gauge couplings is unaffected by the unversality scale
of the soft SUSY-breaking parameters, so in the GUT-less CMSSM, the gaugino masses
become
\begin{equation}
M_a(Q)=\frac{\alpha_a(Q)}{\alpha_a(M_{in})}m_{1/2}.
\label{eq:gauginos}
\end{equation}
Since $\alpha_a(M_{in})$ will be closer to $\alpha_a(Q)$ for $M_{in} <
M_{GUT}$, the low-scale guagino masses will be closer to $m_{1/2}$ as $M_{in}$
is lowered. 

\begin{figure}
\begin{center}
\includegraphics[width=0.40\textwidth,angle=0]{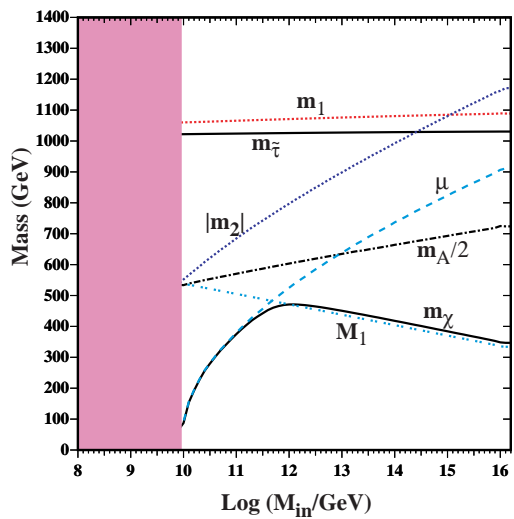}
\end{center}
\caption{Dependence of sparticle mass parameters on the input scale at
which they are assumed to be universal at the point
$(m_{1/2},m_0)=(800,1000)$ GeV with $A_0=0$, $\tan{\beta}=10$, and
$\mu>0$. The pink shading indicates that this point is unphysical for
$M_{in}<10^{10}$ GeV because electroweak symmetry breaking is not obtained.}
\label{fig:masses}     
\end{figure}

In Figure \ref{fig:masses}, we show several low energy sparticle masses, calculated using the full two-loop RGEs, as
functions of the universality scale for $m_{1/2}=800$ GeV and
$m_0=1000$ GeV.  As $M_{in}$ is lowered, $M_1$,
the bino mass, approaches its input scale value of 800 GeV.

Renormalization of the soft SUSY-breaking scalar masses comes from
both gauge and Yukawa interactions, so the running is more
complicated.  To one loop, the effects can be summarized as
\begin{equation}
m_{0_i}^2(Q) = m_0^2 + C_i(Q,M_{in})m_{1/2}^2,
\label{eq:scalars}
\end{equation}
for $Q < M_{in}$, where $m_0$ and $m_{1/2}$ are the universal scalar and
gaugino masses at the input scale. The renormalization coefficients,
$C_i(Q,M_{in})$ vanish as $Q \rightarrow M_{in}$.  As $M_{in}$ is decreased
from $M_{GUT}$, these coefficients diminish, and the low-scale scalar
masses become closer to their input value and therefore less
separated.  

This modification in the running of the soft scalar masses has a
significant impact on the calculated value of the Higgs mass
parameter. At tree-level, the
electroweak vacuum conditions imply
\begin{equation}
\mu^2 = \frac{m_1^2 - m_2^2
\tan^2{\beta}}{\tan^2{\beta}-1}-\frac{M_Z^2}{2},
\label{eq:mu}
\end{equation}
where $m_1$ and $m_2$ are the soft Higgs masses.  As $M_{in}$ is
lowered, these scalar masses become closer to each other, and consequently $\mu^2$
becomes smaller for fixed $m_{1/2}$, $m_0$, and $\tan{\beta}$, as seen
in Figure \ref{fig:masses}. 

Another important consequence of this modification to the running of
the sparticle masses concerns the composition
of the neutralino LSP. In the CMSSM, the neutralino is commonly
bino-like, and $m_{\chi} \sim M_1$.  When $\mu < M_1$, however, the
neutralino is Higgsino-like, and annihilations to vector bosons
are enhanced.  As $M_{in}$ is lowered, we expect that the
LSP becomes more Higgsino-like over all of parameter space. Again
turning to Figure \ref{fig:masses}, we see that for this point in
parameter space, the LSP mass tracks that of the bino for $M_{in}
\geq 10^{12}$ GeV. Below this value, however, $\mu$ becomes much
less than the bino mass, and the LSP eventually becomes Higgsino-like
with $m_{\chi} \sim \mu$.

\section{Evolution of the Relic Density}
\label{sec:evolution}

\begin{figure*}
\begin{center}
\mbox{\includegraphics[height=6.2cm]{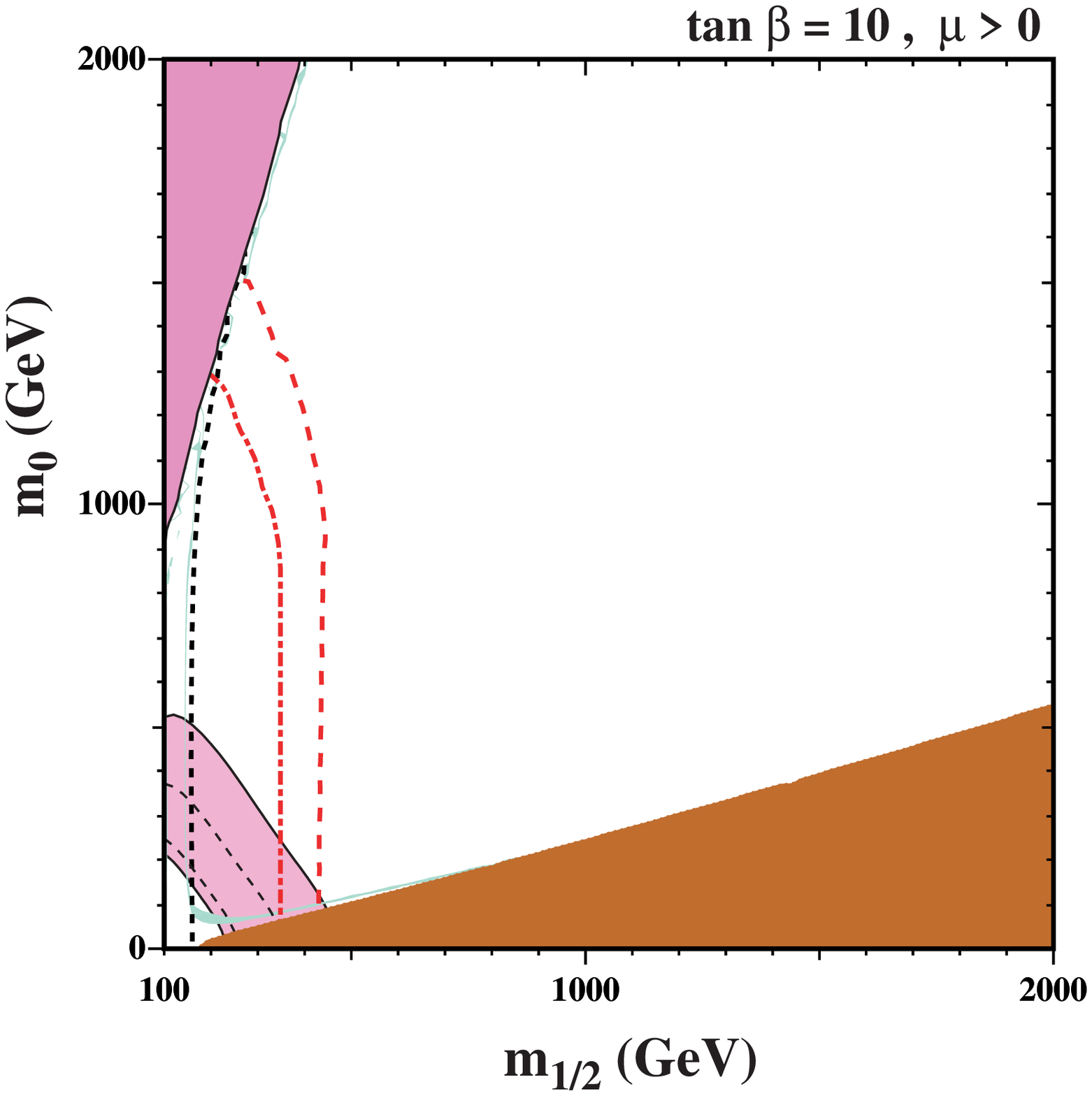}}\qquad
\mbox{\includegraphics[height=6.2cm]{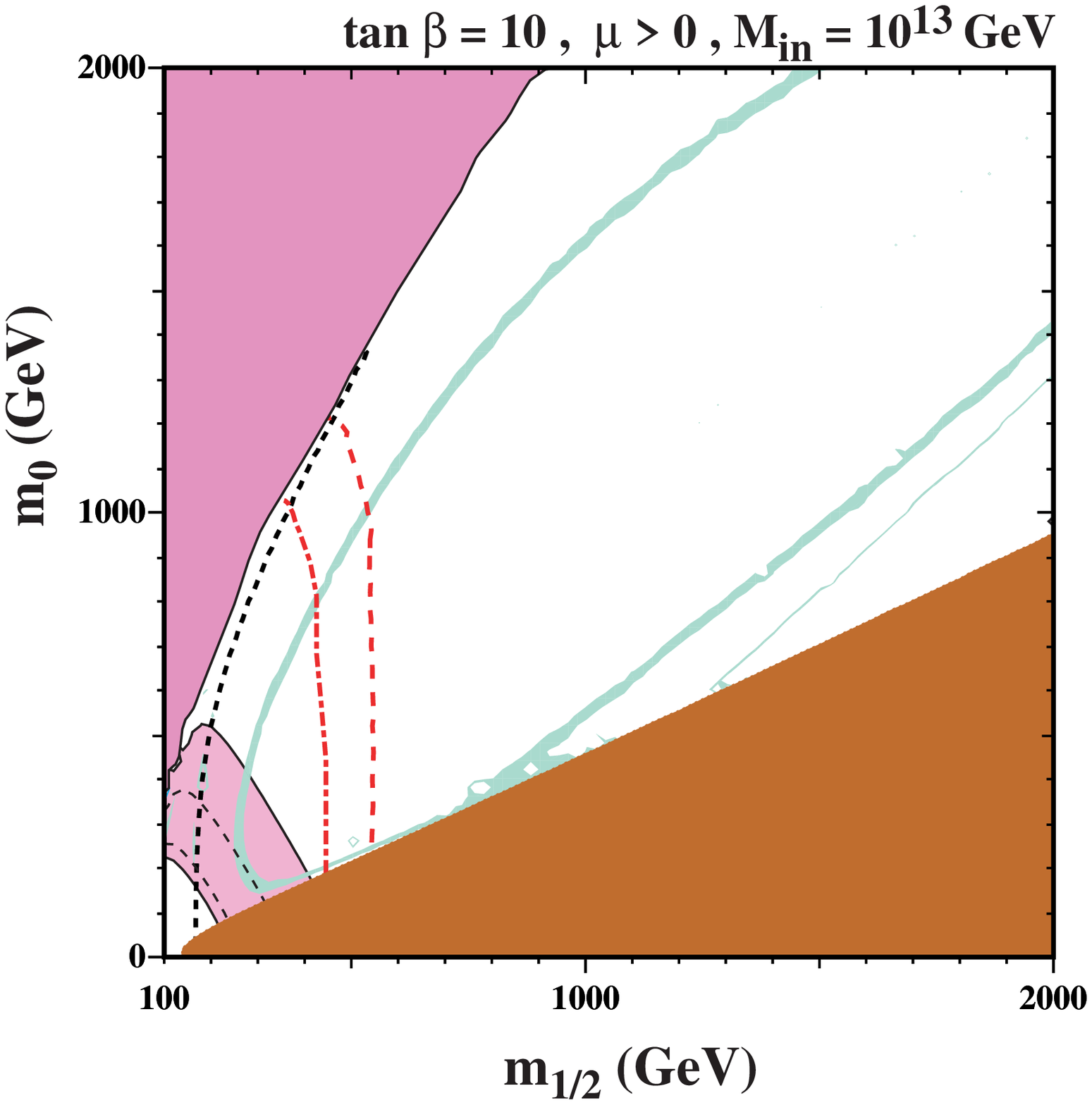}}
\end{center}
\begin{center}
\mbox{\includegraphics[height=6.2cm]{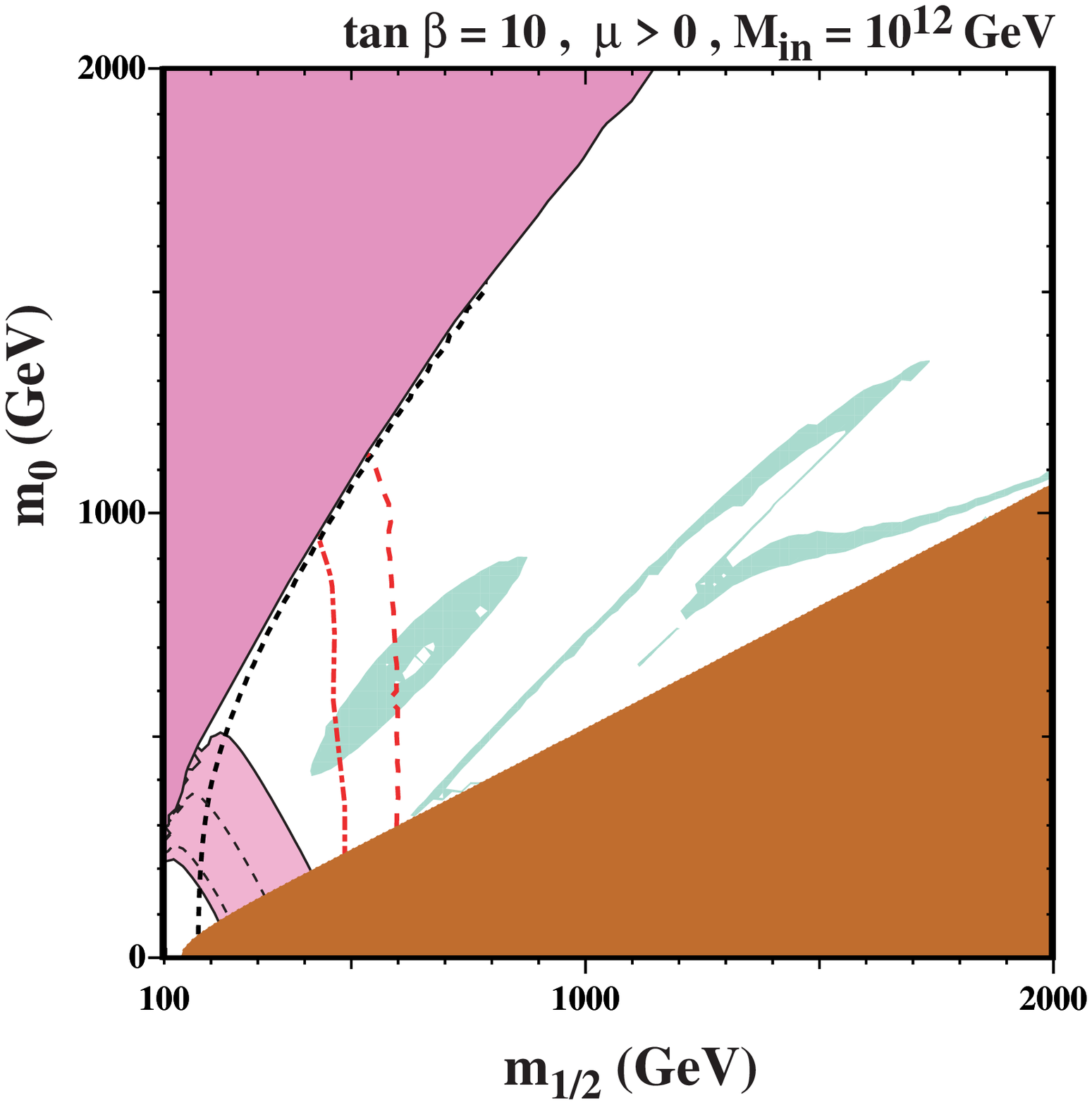}}\qquad
\mbox{\includegraphics[height=6.2cm]{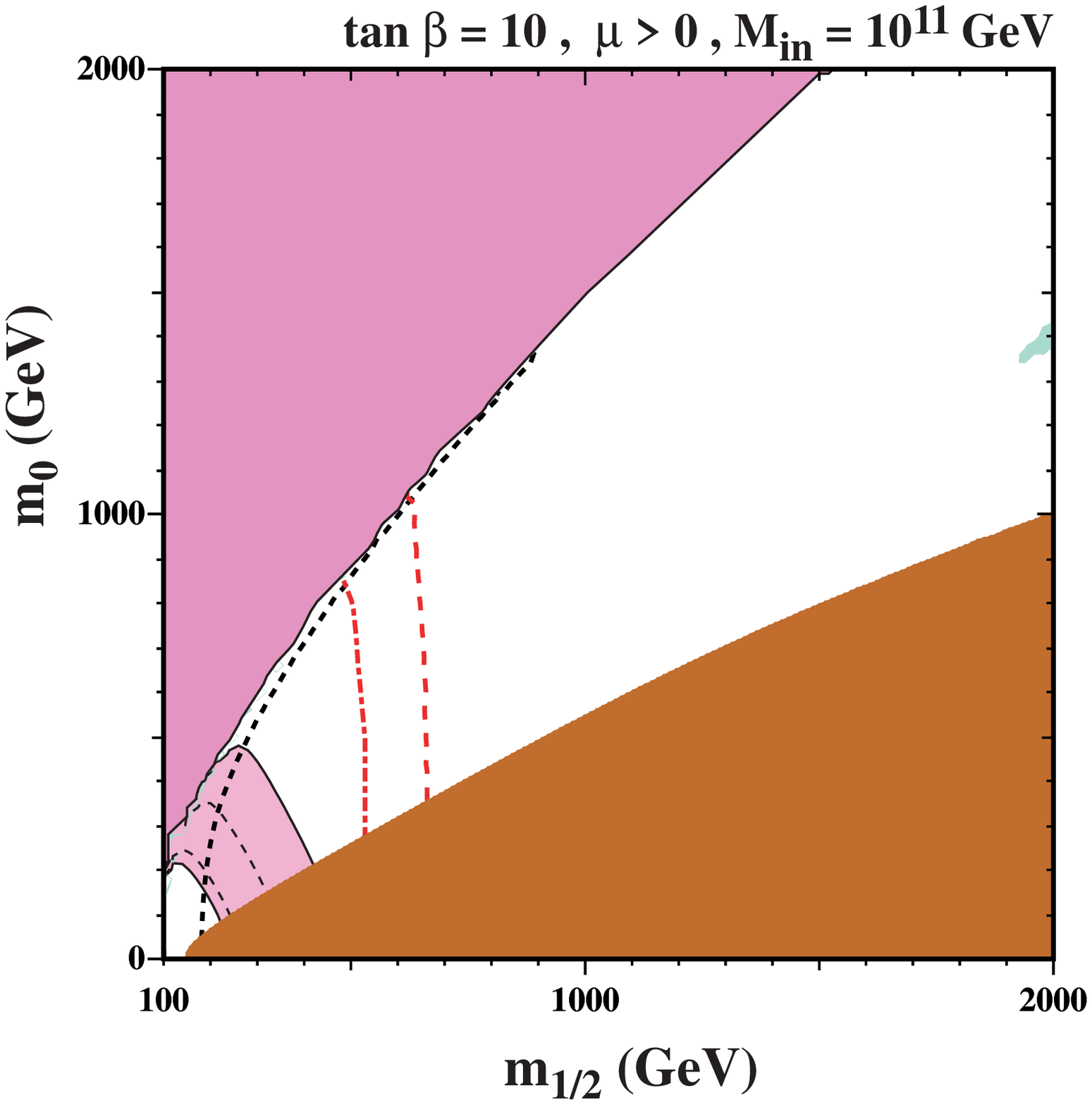}}
\end{center}
\caption{Examples of $(m_{1/2}, m_0)$ planes with $\tan \beta = 10$ and 
$A_0 = 0$ but with  different values of $M_{in}$.
(a) The CMSSM case with $M_{in} = M_{GUT} \sim 2 \times 10^{16}$~GeV, 
(b) $M_{in} = 10^{13}$ GeV, 
(c) $M_{in} = 10^{12}$ GeV and (d) $M_{in} = 10^{11}$ GeV. 
In each panel, we show the region
ruled out because the LSP would be charged (dark red shading), and
that excluded by the electroweak vacuum condition (dark pink
shading). The regions where the relic density falls in the range favored 
by WMAP, $\Omega_{CDM} h^2 =
0.1045^{+0.0072}_{-0.0095}$, have light turquoise shading. In each
panel we also show contours representing the LEP lower limit on the
chargino mass \cite{LEPchargino} (black dashed), a Higgs mass contour
of 114 GeV \cite{LEPhiggs} (red
dashed), and the more exact (and relaxed) Higgs mass bound (red
dot-dashed). The region suggested by $g_{\mu}-2$ at 2-$\sigma$ has
medium (pink) shading with 1-$\sigma$ contours shown as black dashed
lines \cite{gm2}.}
\label{fig:planes}
\end{figure*}

We assume that the neutralino LSP constitutes the cold dark matter in
the universe, and demand that the relic density of neutralinos falls
within the range
\begin{equation}
0.0855 < \Omega_{\chi}h^2 < 0.1189,
\label{eq:wmap}
\end{equation}
in accordance with the WMAP observations \cite{wmap}. In Figure
\ref{fig:planes} we show how
the shape of this constraint changes in the $(m_{1/2},m_0)$ plane as
the universality scale is lowered from the GUT scale. Here we take
$\tan{\beta} =10$, $\mu >0$, $A_0 = 0$ and $m_t = 171.4$ GeV. 

Panel
(a) of Figure \ref{fig:planes} shows the constraints from cosmology
and collider experiments in the
$(m_{1/2},m_0)$ plane in the normal GUT-scale CMSSM. Collider
constraints are described in the figure caption.  The only allowed regions where the relic density of neutralinos falls within the
cosmologically preferred range are the focus point, which borders the
region excluded by the electroweak symmetry breaking condition at
large $m_0$, and the $\chi-\widetilde{\tau}$ coannihilation strip,
which lies along the border of the forbidden $\widetilde{\tau}$-LSP
region. The relic density of neutralinos is much too large to explain
the WMAP measurement over most of the plane.

For $M_{in} = 10^{13}$ GeV, as
shown in Panel (b), the unphysical region where $\mu^2 < 0$ has
encroached further into the plane, as has the forbidden
$\widetilde{\tau}$-LSP region. $\mu$ has become smaller over the
plane, leading to a more Higgsino-like (and lighter) LSP. The
cosmologically preferred strips have pulled away from these excluded
regions, and the two strips on the lower side of the ``C'' shape
indicate the two walls of a rapid annihilation funnel.  Inside these
two walls, $2 m_{\chi} \sim m_{A}$, and neutralino annihilation proceeds extremely
efficiently through the s-channel $A$-pole, resulting in a very low
relic density. 

As the universality scale is lowered to $M_{in} = 10^{12}$ GeV in
Panel (c), the
``C'' has closed into a small island and the rapid annihilation funnel
has broadened and moved to lower $m_{1/2}$. The lower funnel wall has
taken on a peculiar shape, the sharp plunge indicating the $\chi\chi
\rightarrow h+A$ threshold. For this value of $M_{in}$, the LSP is
Higgsino-like and the relic density is below the WMAP range over most
of the plane. 

In Panel (d), where $M_{in}=10^{11}$ GeV, the island has evaporated and
the lower funnel wall has dipped into the $\widetilde{\tau}$-LSP
region such that the relic density of neutralinos is too low to fully
account for the WMAP measurement over
nearly all of the plane. Very low values of $M_{in}$ are therefore
disfavored due to the necessity of a secondary source of astrophysical
cold dark matter in order to explain the measured abundance.

At large $\tan{\beta}$, a similar evolution presents itself, the key
difference being that the rapid annihilation funnel is present already
at $M_{in}=M_{GUT}$. As in the $\tan{\beta}=10$ case, the upper funnel
wall and what had been the focus point region approach, merge, and
eventually evaporate, at which point the lower funnel wall falls into
the $\widetilde{\tau}$-LSP region so that the relic density of
neutralinos is too low over the entire plane when $M_{in} \sim 10^{11}$ GeV.

\section{Direct Detection}
\label{sec:detection}

\begin{figure*}
\begin{center}
\mbox{\includegraphics[height=6.2cm]{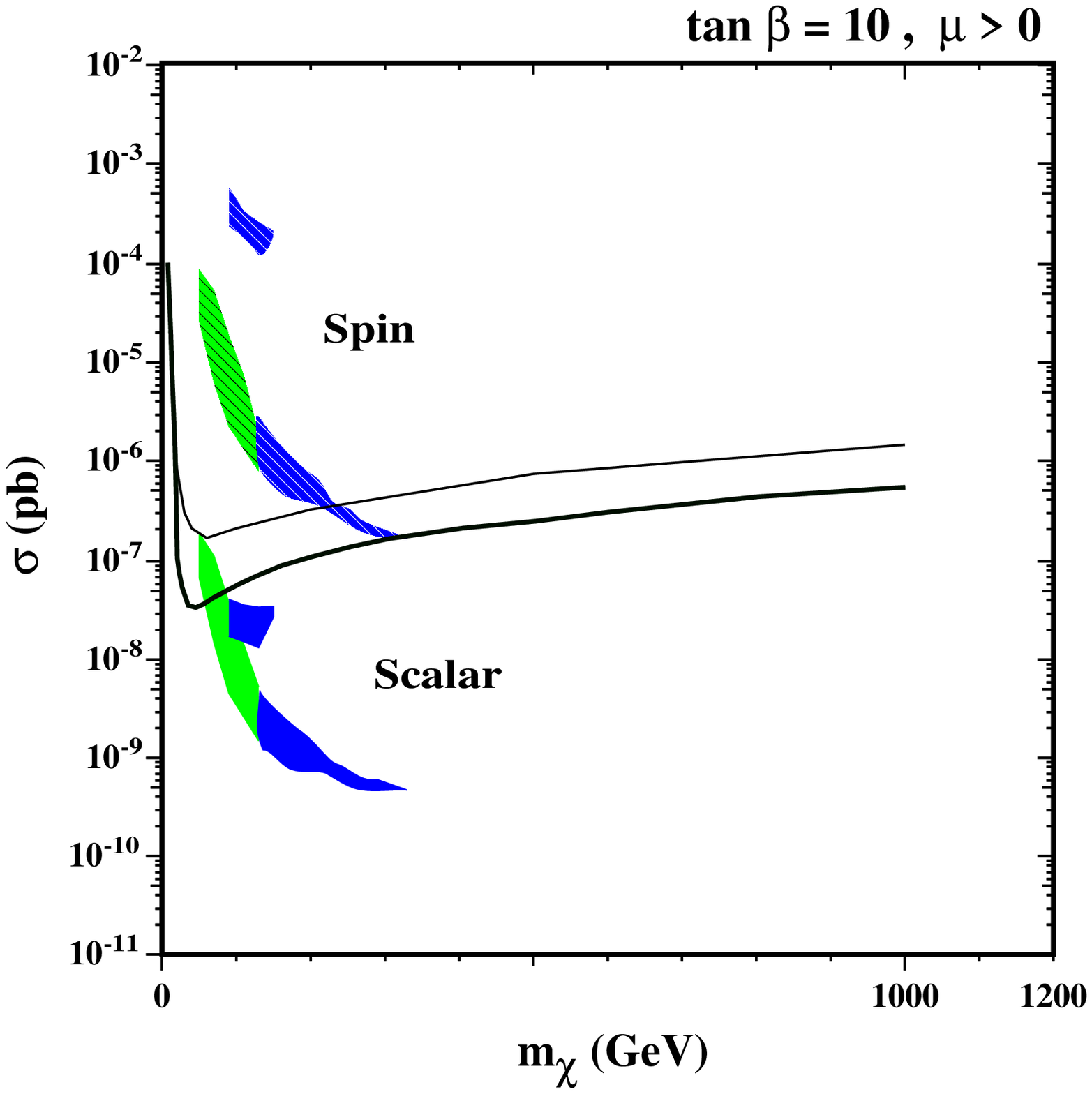}}\qquad
\mbox{\includegraphics[height=6.2cm]{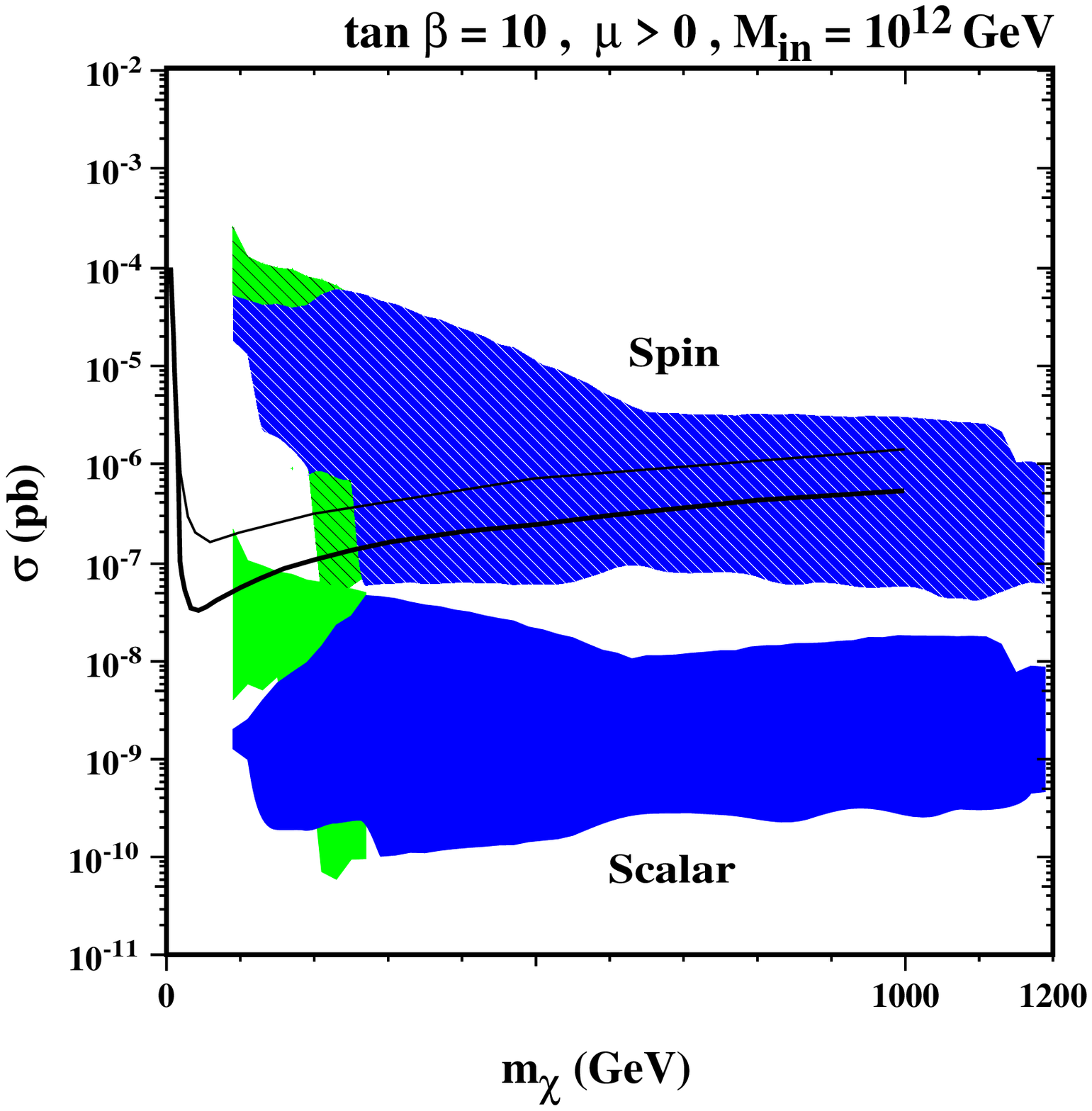}}
\end{center}
\caption{Neutralino-nucleon spin-dependent (hatched) and scalar
(solid) cross sections as functions of the
neutralino mass for $\tan{\beta}=10$ and $A_0=0$ with $M_{in}=M_{GUT}$
in Panel (a) and $M_{in}=10^{12}$ GeV in Panel (b).}
\label{fig:scattering}
\end{figure*}

Many direct searches for dark matter particles such as CDMS and
XENON10 look for signatures of weakly-interacting massive particles
(WIMPs) scattering on nuclei. The neutralino-nucleon scattering cross
section can be broken into spin-dependent and spin-independent
(scalar) parts. In Figure \ref{fig:scattering}, we plot the
neutralino-nucleon elastic scattering cross sections as functions of
neutralino mass for regions in our parameter space that pass all constraints (blue) and that fail only the relaxed
LEP Higgs constraint\footnote{See \cite{eos2} for details.} (green) with $M_{in}=M_{GUT}$ in Panel (a) and
$M_{in}=10^{12}$ GeV in Panel (b). If the relic density is below the
central value, $\Omega_{\chi}h^2 = 0.1045$, we plot the cross section
scaled by the ratio of the relic density of neutralinos to the
measured value such that our cross sections can be compared with the
exclusion limits from direct detection experiments. In particular, we
show the limits from CDMS II and XENON10 for the scalar part of the neutralino-nucleon cross section
\cite{cdms,xenon10}.

In Panel (a), where $M_{in}=M_{GUT}$, the regions that pass all
constraints are separated into a longer strip of lower cross sections
and a smaller island of larger cross sections. The longer strip
corresponds to points in the $\chi$-$\widetilde{\tau}$ coannihilation
region, while the smaller island corresponds to the focus point
region.  We note that this island would extend to larger $m_{\chi}$
had we considered values of $m_0 > 2000$ GeV.

When $M_{in}=10^{12}$ GeV, as shown in Panel (b), the most striking
difference is that for a given value of $m_{\chi}$, there are a wide
range of potential cross sections. This is due to the fact that the
relic density of neutralinos is below the WMAP range over most of the
plane, so cross sections are scaled accordingly. The upper boundaries
of the regions pictured in Panel (b) come from points in the
$(m_{1/2},m_0)$ plane where the relic density is the largest and $m_0$
is the lowest, so one can map the upper edges in Panel (b) to
cosmologically preferred regions in Panel (c) of Figure
\ref{fig:planes}.

The next generation of direct detection experiments will probe much of
the range of scalar cross sections pictured here.  SuperCDMS Phase A
with seven towers deployed is expected to be sensitive to scalar cross
sections as low as $10^{-9}$ pb at $m_{\chi} =100$ GeV
\cite{supercdms}. Detectors that will use liquid nobles such as Xenon
or Argon expect sensitivities as low as $10^{-10}$ pb for one year
of operation of a one ton detector \cite{ardm}. We look forward to increased
sensitivities to neutralino-nucleon cross sections as a useful
complement to collider searches.

\section{Conclusions}
\label{sec:conclusions}

Lowering the unification scale of the soft SUSY-breaking mass
parameters significantly changes the appearance of the constraint on the relic density of
neutralinos. Intermediate universality results in the merging and evaporation of the focus point and rapid annihilation funnel
regions. For some critical
value of $M_{in}$, dependent on $\tan{\beta}$ and other factors, the
relic density of neutralinos becomes too low over all or nearly all of
the $(m_{1/2},m_0)$ plane to explain the
measured relic density of cold dark matter. The dark matter
phenomenology of GUT-less SUSY represents a substantial departure
from that of the standard GUT-scale CMSSM.

\section*{Acknowledgments}
\noindent
The work of P.S. was supported in part
by DOE grant DE--FG02--94ER--40823.

%
%

\end{document}